# Vortexes as systems specific to the Acoustic World

Ion Simaciu[a,*], Viorel Drafta[b], Zoltan Borsos[c,**] and Gheorghe Dumitrescu[d]

[a] Retired lecturer, Petroleum-Gas University of Ploiești, Ploiești 100680, Romania

[b] Independent researcher

[c] Petroleum-Gas University of Ploiești, Ploiești 100680, Romania

[d] Retired professor, High School Toma N. Socolescu, Ploiești, Romania

E-mail: [*] isimaciu@yahoo.com; [**] borzolh@upg-ploiesti.ro

Keyword: Acoustic World, vortex, hydrodynamic Maxwell equations, angular momentum, vortex lens

Abstract:

In this paper we study the properties of vortexes, as systems specific to the Acoustic World, using both hydrodynamic theory and the corresponding hydrodynamic Maxwell equations. According to this study, it follows that the vortex behaves like an acoustic dipole that has intrinsic/internal angular momentum. The system of two identical vortices also has orbital angular momentum and behaves, at distances much greater than the distance between the axes of the vortices, as a single vortex. With the help of Maxwell's hydrodynamic equations for the vortex we deduced the force between two vortices and obtained the expression of the equivalent mass of the vortex and the permittivity of the electroacoustic field. We also obtained and interpreted the expression for the energy density of the acoustic field. The density and pressure variations induced by the vortex cause the change in the propagation speed of the acoustic waves and the acoustic lensing property of the vortex.

## 1. Introduction

An important stage in the definition and study of the Acoustic World (AW) is the identification of specific systems [1, 2]. These systems are constituted for AW as the elementary "particles" from which, through interaction, more complex systems are formed: the acoustic "atom", the bubble clusters (the system of $N$ "particles" and/or "atoms"), etc. To date we have specifically identified and studied the following systems: acoustic waves and acoustic wave packets [1, 3-6], acoustic bubbles [7-14] and bubble clusters [15-19]. For these specific systems we have identified and theoretically studied the following properties: The dual (particle-wave) properties of the acoustic wave and the wave packet [1-4], the lens and the dumb hole corresponding to an acoustic wave packet [5, 6], the interaction between two pulsating bubbles (secondary Bjerknes forces) and the corresponding acoustic charge [7-9, 15], average acoustic charge in the acoustic radiation background [7-9], gravitational acoustic interaction [12, 13], properties (resonance and acoustic Mach's principle, internal pressure, internal energy, cluster acoustic temperature) of the system of identical pulsating bubbles [16-20], the acoustic lens corresponding to a pulsating bubble [10, 11] and the acoustic lens and acoustic hole corresponding to a cluster of pulsating bubbles [18, 19].

In this paper we study the properties of another system specific to the AW, the vortex. This phenomenon and at the same time physical system has been studied both theoretically [21, 22] and experimentally [23, 24] because it has numerous effects on fluid dynamics in turbulent flow [25] and bodies moving in fluids and therefore technological implications [26].

In the second part we study the types of vortices and their properties. In the third part we study the interaction between two vortices using Maxwell's hydrodynamic equations. In the fourth part we study the behavior as an acoustic lens of the vortex. The last part is dedicated to the conclusions.

## 2. Vortexes

### 2.1. Vortex, definition and classification

The vortex is characterized by the physical quantity vorticity, which characterizes the property of rotating the fluid flow around the axis. Mathematically, vorticity is described by an axial vector, which characterizes the rotation at a point in the fluid. If we describe the velocity distribution by the velocity field $\vec{v}(\vec{r}), \vec{r} = x\hat{x} + y\hat{y}$ (with $r = \sqrt{x^2 + y^2}$ the distance from the axis) or $\vec{r} = (r\cos\theta)\hat{x} + (r\sin\theta)\hat{y}$ in cylindrical coordinates in the plane $Oxy$), the connection between the vorticity and the velocity field is [22]

$$\vec{\omega} = \nabla \times \vec{v}. \tag{1}$$

There are two types of vortices: the rigid-body vortex and the irrotational vortex. The rigid body-vortex or the rotational vortex is the vortex in which the fluid rotates as a rigid body. For this type of vortex, the angular velocity $\vec{\Omega}$ does not depend on the position (distance from the axis $r$) and between the vorticity vector, the velocity vector and the angular velocity vector there are the relations:

$$\vec{\omega} = \nabla \times \vec{v} = 2\vec{\Omega}, \quad \vec{v} = \vec{\Omega} \times \vec{r}, \vec{v} = \Omega r \hat{\theta}. \tag{2}$$

The irrotational vortex is the vortex in which the velocity of the fluid decreases inversely proportional to the distance to the axis:

$$\vec{\Omega} = \frac{\Gamma}{2\pi r^2}\hat{z}, \ \vec{v} = \vec{\Omega} \times \vec{r} = \frac{-\Gamma y}{2\pi r^2}\hat{x} + \frac{\Gamma x}{2\pi r^2}\hat{y}, \ \vec{\omega} = \nabla \times \vec{v} = 0 \tag{3}$$

that is, in polar coordinates, $\vec{v} = \Gamma\hat{\theta}/(2\pi r)$ and $v_x = -\Gamma\sin\theta\hat{x}/(2\pi r), v_y = \Gamma\cos\theta\hat{y}/(2\pi r)$. In Eq. (3), $\Gamma$ is the circulation of the flow velocity on a closed curve $C$ that includes the origin of the reference system (intersection of the plane $Oxy$ with the axis of rotation $Oz$)

$$\Gamma = \oint_C \vec{v} d\vec{l} = \text{const.}, \tag{4}$$

with the elementary length vector $d\vec{l}$ parallel to the tangent vector $\vec{v}$, to the curve $C$, at each point. This vortex is characterized by the non-physical property that the speed tends to infinity when the distance tends to zero, $r \to 0$. This vortex is known as the point or line vortex.

A vortex model closer to real (experimental) situations is the core vortex. This vortex, for a viscous fluid, was studied by William John Macquorn Rankine [27]. This type of vortex has a core of radius $D$ which rotates as a rigid with angular velocity $\vec{\Omega}$ and where, for distances from the axis $r \leq D$, the fluid velocity is given by Eq. (2). For larger distances $r > D$, the vortex is irrotational having zero vorticity and the relation between the angular velocity and the fluid velocity is given by Eq. (3).

In specialized papers [22, 28-31], other types of vortices are also described: the spherical vortex (Hill vortex), the ring vortex (toroidal vortex, circular vortex ring), Burgers vortex (an axisymmetric viscous vortex) and the Lamb–Oseen vortex.

Of particular importance for the modeling of the Acoustic World is the theoretical and experimental study of vortices in quantum superfluids and superconductors [32-36].

## 2.2. Two vortex system

In the paper "Elementary Fluid Mechanics" [22], the system of two point vortices is studied, using the complex potentials $F(z) = [k/(2\pi i)] \ln z$ with $z = x + iy, \bar{z} = x - iy$. The potential of the two-vortex system is

$$F(z) = F_1(z) + F_1(z) = \frac{k_1}{2\pi i} \ln(z - z_1) + \frac{k_2}{2\pi i} \ln(z - z_2). \tag{5}$$

With this potential, the velocities of the two vortices are:

$$\frac{d\bar{z}_1}{dt} = \dot{\bar{z}}_1 = \left.\frac{dF_2(z)}{dz}\right|_{z-z_1} = \frac{k_2}{2\pi i} \frac{1}{z_1 - z_2}, \frac{d\bar{z}_2}{dt} = \dot{\bar{z}}_2 = \left.\frac{dF_1(z)}{dz}\right|_{z-z2} = \frac{k_1}{2\pi i} \frac{1}{z_2 - z_1} \tag{6}$$

or

$$\dot{\bar{z}}_1 = \dot{x}_1 - i\dot{y}_1 = \frac{k_2}{2\pi} \frac{(y_1 - y_2) - i(x_1 - x_2)}{(x_1 - x_2)^2 + (y_1 - y_2)^2},$$

$$\dot{\bar{z}}_2 = \dot{x}_2 - i\dot{y}_2 = \frac{k_1}{2\pi} \frac{(y_2 - y_1) - i(x_2 - x_1)}{(x_2 - x_1)^2 + (y_2 - y_1)^2}. \tag{7}$$

The distance between the two vortices, $d^2 = (x_1 - x_2)^2 + (y_1 - y_2)^2$, is constant and the center of vorticity [22] is characterized by the complex quantity

$$z_c = \frac{k_1 z_1 + z_2 k_2}{k_1 + k_2}. \tag{8}$$

If we consider the origin of the reference system to be in $z_c$ (i.e. $z_c = 0$), the relation follows

$$k_1 z_1 = -z_2 k_2. \tag{9}$$

With the help of relations (8, 9), we define the complex distances between the center of each vortex and the center of vorticity:

$$z_1 - z_c = \frac{k_1(z_1 - z_2)}{k_1 + k_2}, \quad z_2 - z_c = \frac{k_2(z_2 - z_1)}{k_1 + k_2}. \tag{10}$$

The distances between the centers of the vortices and the center of vorticity, obtained using Eqs. (10), are:

$$d_{x1} = \frac{k_2(x_1 - x_2)}{k_1 + k_2}, d_{y1} = \frac{k_2(y_1 - y_2)}{k_1 + k_2}, d_1 = \frac{k_2 d}{k_1 + k_2};$$

$$d_{x2} = \frac{k_1(x_2 - x_1)}{k_1 + k_2}, d_{y1} = \frac{k_1(y_2 - y_2)}{k_1 + k_2}, d_2 = \frac{k_1 d}{k_1 + k_2}, d^2 = d_1^2 + d_2^2. \tag{11}$$

With these quantities, the expressions of the velocities of the two vortices, given by Eqs. (7, 10), become:

$$\upsilon_{12} = \sqrt{\dot{x}_1^2 + \dot{y}_1^2} = \frac{k_2}{2\pi d} = \frac{k_2^2}{2\pi(k_1 + k_2)d_1}, \quad \upsilon_{21} = \sqrt{\dot{x}_2^2 + \dot{y}_2^2} = \frac{k_1}{2\pi d} = \frac{k_1^2}{2\pi(k_1 + k_2)d_2}. \tag{12}$$

In polar coordinates, the positions of the two vortices are: $z_j = r_j \exp(i\theta_j), j = 1, 2$ and the angular velocities are:

$$\frac{d\theta_1}{dt} = \dot{\theta}_1 = \frac{\upsilon_{12}}{d_1} = \frac{k_2}{2\pi d_1 d} = \frac{k_2^2}{2\pi(k_1+k_2)d_1^2} = \frac{k_1+k_2}{2\pi d^2},$$
$$\frac{d\theta_2}{dt} = \dot{\theta}_2 = \frac{\upsilon_{21}}{d_2} = \frac{k_1}{2\pi d_2 d} = \frac{k_1^2}{2\pi(k_1+k_2)d_2^2} = \frac{k_1+k_2}{2\pi d^2} = \dot{\theta}_1 = \dot{\theta}.$$
(13)

It follows that the two vortices form a more complex system that behaves as a single vortex at distances greater than the distance between their centers ($r > d$). If the vortices are different, $k_1 \neq k_2$, but have different orientations, $k_1 > 0, k_2 < 0$, the resulting vortex will have angular velocity $\dot{\theta} = [k_1 - |k_2|]/(2\pi d^2)$. If the vortices are identical, $k_1 = k_2 = k$, and have the same vorticity orientation, the resulting vortex has angular velocity $\dot{\theta} = k/(\pi d^2)$. If the vorticities are opposite and equal in value, $k_1 = k, k_2 = -k$, the angular velocity of the system is zero and the system performs a translational movement, in the direction perpendicular to the line joining their centers, with linear velocity $\upsilon = k/(2\pi d)$ [22]. The solutions given by this formalism do not tell us what is the inertial mass of the vortex and therefore what is the expression of the force between the two vortices.

## 3. Hydrodynamic Maxwell equations for the vortex

### 3.1. Hydrodynamic Maxwell equations

In the specialized scientific papers [37, 38, 15], starting from the analogy between electromagnetism and fluid mechanics, the Maxwell hydrodynamic equations were deduced, which, for the vortex, have the form:

$$\nabla \cdot \vec{E}_\upsilon = 4\pi \rho_\upsilon, \tag{14}$$

$$\nabla \cdot \vec{H}_\upsilon = 0, \tag{15}$$

$$\nabla \times \vec{E}_\upsilon + \frac{\partial \vec{H}_\upsilon}{\partial t} = 0, \tag{16}$$

$$u^2 \left(\nabla \times \vec{H}_\upsilon\right) - \frac{\partial \vec{E}_\upsilon}{\partial t} = \vec{J}_\upsilon \equiv 4\pi \vec{j}_\upsilon, \tag{17}$$

with the following defining relations for the physical quantities involved: the field vectors

$$\vec{E}_\upsilon \equiv -\frac{\partial \vec{\upsilon}_\upsilon}{\partial t} - \nabla h_\upsilon, \quad \vec{H}_\upsilon \equiv \vec{\omega} = \nabla \times \vec{\upsilon}_\upsilon, \quad h_\upsilon = \frac{\upsilon_\upsilon^2}{2}; \tag{18}$$

the acoustic charge density $\rho_\upsilon$ and the acoustic current density, $\vec{J}_\upsilon \equiv 4\pi \vec{j}_\upsilon = 4\pi \rho_\upsilon \vec{\upsilon}$,

$$4\pi \rho_\upsilon = -\frac{\partial (\nabla \vec{\upsilon}_\upsilon)}{\partial t} - \Delta h_\upsilon, \tag{19}$$

$$\vec{J}_\upsilon \equiv 4\pi \vec{j}_\upsilon = -\frac{\partial^2 \vec{\upsilon}_\upsilon}{\partial t^2} + \nabla\left(\frac{\partial h_\upsilon}{\partial t}\right) + u^2 \left(\nabla \times (\nabla \times \vec{\upsilon}_\upsilon)\right). \tag{20}$$

These hydrodynamic equations correspond to the energy densities, $w = w_E + w_H$

$$w = \frac{\vec{E}_\upsilon^2}{8\pi} + \frac{\vec{H}_\upsilon^2}{8\pi} \rightarrow w = \frac{\varepsilon_\upsilon \vec{E}_\upsilon^2}{2} + \frac{\mu_\upsilon \vec{H}_\upsilon^2}{2}. \tag{21}$$

written in the Gussian and MKSA system of units [40]. We consider a vortex with a core and the fluid is considered ideal with zero viscosity. In the nucleus of radius $D$ ($r \leq D$), according to relation (2), the velocities are independent of time:

$$\vec{\upsilon}_{\upsilon i} = \vec{\Omega} \times \vec{r} = (0, \Omega r, 0), \vec{\upsilon}_{\upsilon i} = \Omega r \hat{\theta}; \vec{\omega}_i = \nabla \times \vec{\upsilon}_{\upsilon i} = 2\vec{\Omega} = (0, 0, 2\Omega); \qquad (22)$$

and outside ($r > D$), are:

$$\vec{\upsilon}_{\upsilon e} = (0, \frac{\Omega D^2}{r}, 0), \vec{\upsilon}_{\upsilon e} = \frac{\Omega D^2}{r} \hat{\theta}; \vec{\omega}_e = \nabla \times \vec{\upsilon}_{\upsilon e} = 0. \qquad (23)$$

With definitions (18) and (22), we obtain for the core vortex: the interior field vectors:

$$\vec{E}_{\upsilon i} \equiv -\nabla h_\upsilon = -\upsilon_\upsilon (\nabla \upsilon_\upsilon) = -\Omega^2 \vec{r} = (-\Omega^2 r)\hat{r}, \ \vec{H}_{\upsilon i} = 2\vec{\Omega} = 2\Omega \hat{z} \qquad (24)$$

and the exterior field vectors

$$\vec{E}_{\upsilon e} \equiv -\nabla h_{\upsilon e} = -\upsilon_{\upsilon e}(\nabla \upsilon_{\upsilon e}) = \frac{\Omega^2 D^4}{r^4}\vec{r}, \ \vec{H}_{\upsilon e} = 0. \qquad (25)$$

The units of measurement, in the International System (SI), of the intensity of the acoustic field are those of an acceleration, $\langle E_{\upsilon e} \rangle_{SI} = \text{m/s}^2$, and those of the acoustic induction are those of the angular velocity, $\langle H_\upsilon \rangle_{SI} = 1/\text{s}$.

With definitions (19) and (22), the acoustic charge density $\rho_a$ and the acoustic current density $\vec{J}_a$, become:

$$\rho_{\upsilon i} = \frac{-\Delta h_{\upsilon i}}{4\pi} = \frac{-\Delta(\upsilon_{\upsilon i}^2)}{8\pi} = \frac{-\Omega^2}{2\pi}, \rho_{\upsilon e} = \frac{-\Delta(\upsilon_{\upsilon e}^2)}{8\pi} = \frac{-\Omega^2 D^4}{2\pi r^4}; \qquad (26)$$

$$\vec{J}_{\upsilon i} \equiv 4\pi \vec{j}_{\upsilon i} = u^2 \left(\nabla \times (2\vec{\Omega})\right) = 0, \ \vec{j}_{\upsilon i} = 0. \qquad (27)$$

The acoustic charge of the vortex, per unit length, is calculated by integration: $e_\upsilon = \int_0^r \rho_\upsilon 2\pi r dr$. For the case when $r \ll D$, the internal charge that generates the acoustic intensity inside results:

$$e_{\upsilon i} = \int_0^r \rho_{\upsilon i} 2\pi r dr = \int_0^r \left(\frac{-\Omega^2}{2\pi}\right) 2\pi r dr = \frac{-\Omega^2 r^2}{2}, e_{\upsilon iD} = e_{\upsilon i}(r = D) = \frac{-\Omega^2 D^2}{2} \qquad (28)$$

With this charge, the expression for the internal field strength (24) becomes

$$\vec{E}_{\upsilon i} = (-\Omega^2 r)\hat{r} = \frac{2e_{\upsilon i}\hat{r}}{r}. \qquad (29)$$

In Eq. (29) of the field intensity expressed by the charge that determines it, $e_{\upsilon i}$, the factor 2 appears, because this field has cylindrical symmetry and we used Eq. (14) written for a spherical symmetry. For cylindrical symmetry, $\rho_{\upsilon i} = -\Omega^2/\pi$ and $e_{\upsilon i} = -\Omega^2 r^2$ and $\vec{E}_{\upsilon i} = e_{\upsilon i}\hat{r}/r$. The force between the two vortices is determined by both the field strength $\vec{E}_{\upsilon i}$ and the field induction $\vec{H}_{\upsilon i}$:

$$\vec{f}_{\upsilon i} \equiv e_{\upsilon iD} \vec{E}_{\upsilon i} + e_{\upsilon iD}\left(\vec{\upsilon}_{\upsilon i} \times \vec{H}_{\upsilon i}\right) = (-\Omega^2 D^2)(-\Omega^2 r)\hat{r} + (-\Omega^2 D^2)\left[(\Omega r \hat{\theta}) \times (2\Omega \hat{z})\right]$$
$$= \Omega^4 D^2 r \hat{r} - 2\Omega^4 D^2 r \hat{r} = -\Omega^4 D^2 r \hat{r} = \frac{-e_{\upsilon iD} e_{\upsilon i} \hat{r}}{r}, \langle \vec{f}_{\upsilon i} \rangle_{SI} = \frac{\text{m}^3}{\text{s}^4}. \qquad (30)$$

and it is attractive because the magnetic acoustic component is attractive and twice as large as the electrical component which is repulsive. For the force per unit length to be the Newtonian force $\langle \vec{f}_{\upsilon N} \rangle_{SI} = \text{kg/s}^2$, we must multiply the expression in Eq. (30) with a constant having the physical dimensions $\langle \alpha_N \rangle_{SI} = \text{kgs}^2/\text{m}^3$, $\alpha_N = \alpha \rho_0/\Omega^2$:

$$\vec{f}_{\upsilon i N} = \alpha_N \vec{f}_{\upsilon i} = \frac{\alpha \rho_0}{\Omega^2}(-\Omega^4 D^2 r \hat{r}) = -\alpha \rho_0 D^2 \Omega^2 r \hat{r} = \frac{-\alpha(\rho_0 \pi D^2 l)(2\Omega^2 r)\hat{r}}{2\pi l} \frac{1}{r} = \frac{-\alpha m \vec{a}_{cf}}{2\pi l}, \qquad (31)$$

with $\vec{a}_{cf} = 2v_{vi}^2/r = 2\Omega^2 r$ centrifugal acceleration, $m = \rho_0 \pi D^2 l$ vortex core mass of length $l$ and $\alpha$ numerical/ dimensionless constant. According to section 3, the two vortices of the same direction rotate around the center of mass and so $d_1 = d_2 = r/2$. The constant $\alpha = 2\pi$ is deduced from the condition, $\vec{f}_{viN} = -m\vec{a}_{cf}/l$. In this case, the equivalent mass of the vortex is the mass of the vortex core.

For the case when $r \gg 2D$, the charge of the vortex that generates the intensity of the electroacoustic field is:

$$e_{ve}(r) = \int_0^D \rho_{vi} 2\pi r dr + \int_D^r \rho_{ve} 2\pi r dr = \int_0^D \left(-\frac{\Omega^2}{2\pi}\right) 2\pi r dr + \int_D^r \left(-\frac{\Omega^2 D^4}{2\pi r^4}\right) 2\pi r dr = -\Omega^2 D^2 \left(1 - \frac{D^2}{2r^2}\right) \quad (32)$$

and the total charge of a sample vortex placed outside another identical vortex is

$$e_{vet} = \int_0^D \rho_{vi} 2\pi r dr + \int_D^\infty \rho_{ve} 2\pi r dr = \int_0^D \left(-\frac{\Omega^2}{2\pi}\right) 2\pi r dr + \int_D^\infty \left(-\frac{\Omega^2 D^4}{2\pi r^4}\right) 2\pi r dr = -\Omega^2 D^2. \quad (33)$$

With this charge (32), the intensity of the electroacoustic field (25) is written

$$\vec{E}_{ve} = \frac{\Omega^2 D^4}{r^4}\vec{r} = \frac{e_{ve}(r)D^2}{r^4\left(1 - \frac{D^2}{2r^2}\right)}\vec{r}. \quad (34)$$

The force between two vortices placed at distance $r \gg 2D$, using Eqs. (25) and (32), is

$$\vec{f}_{vet} \equiv e_{vet}\vec{E}_{ve} + e_{vet}\left(\vec{v}_{ve} \times \vec{H}_{ve}\right) = e_{vet}\vec{E}_{ve} = \left(-\Omega^2 D^2\right)\left(\frac{\Omega^2 D^4 \hat{r}}{r^3}\right) = \frac{-\Omega^4 D^6}{r^3}\hat{r}. \quad (35)$$

This force can be written as a function of the expression of the charges (32) and (33) and the moment of the acoustic charge $d_{ve} = e_{ve} D$:

$$\vec{f}_{vet} = \frac{-\left(\Omega^2 D^2\right)^2 D^2 \hat{r}}{r^3} = \frac{-e_{vet}^2 D^2 \hat{r}}{r^3} = \frac{-d_{vet}^2 \hat{r}}{r^3}; \vec{f}_{vet} = \frac{-e_{vet} e_{ve}(r) D^2 \hat{r}}{r^3 \left(1 - \frac{D^2}{2r^2}\right)} = \frac{-d_{vet} d_{ve}(r) \hat{r}}{r^3 \left(1 - \frac{D^2}{2r^2}\right)}. \quad (36)$$

and so it is an acoustic dipole force and is analogous to the electrostatic dipole force [40].

To obtain the Newtonian interaction force of the vortices, per unit length, we must multiply this expression by the same constant $\langle \alpha_N \rangle_{SI} = \text{kgs}^2/\text{m}^3$, $\alpha_N = \alpha \rho_0 / \Omega^2$

$$\vec{f}_{veN} = \alpha_N \vec{f}_{vet} = \frac{-\alpha \rho_0 \Omega^2 D^6 \hat{r}}{r^3} = \frac{\alpha\left(-\rho_0 \pi D^2 l\right)\hat{r}}{2\pi l}\left(\frac{2\Omega^2 D^4}{r^3}\right) = \frac{-\alpha m \vec{a}_{cfe}}{2\pi l}. \quad (37)$$

The expression of this force becomes $\vec{f}_{viN} = -m\vec{a}_{cfe}/l$ if $\alpha = 2\pi$, that is, the numerical constant is of the same magnitude as in the case of the interior of the vortex core. With this constant, the acoustic charge in the Gaussian system is $e_{vetN}^2 = 2\pi\rho_0 \Omega^2 D^4$. And in this case, the equivalent mass of the vortex is the mass of the vortex core $m = \rho_0 \pi D^2 l$.

By analogy with the Maxwell equations for the oscillating bubble that behaves as an acoustic monopole with spherical symmetry, the constant $\langle \alpha_N \rangle_{SI} = \text{kgs}^2/\text{m}^3$, $\alpha_N = 2\pi \rho_0/\Omega^2$ is expressible as a function of the acoustic permittivity $\varepsilon_v = \rho_0/\Omega^2$ for the vortex, $\alpha_N = 2\pi\varepsilon_v$, which depends on the parameters of the fluid of undisturbed density $\rho_0$ and subject to periodic motion (rotational motion with angular velocity $\vec{\Omega}$). Acoustic permeability has the expression $\mu_v = 1/(\varepsilon_v u_0^2) = \Omega^2/(\rho_0 u_0^2)$, $\langle \mu_v \rangle_{SI} = \text{m/kg}$.

To highlight the existence of acoustic permittivity $\varepsilon_v = \rho_0/\Omega^2$, we write the Newtonian force per unit length (37) as a function of the acoustic load in the MKSA/ SI system (i.e., Eq. (14) becomes $\nabla \cdot \vec{E}_v = \rho_v / \varepsilon_v$) [40]:

$$\vec{f}_{veN} = \frac{-\alpha \rho_0 \Omega^2 D^6 \hat{r}}{r^3} = \frac{-q^2 D^2 \hat{r}}{2\pi \varepsilon_v r^3} \tag{38}$$

or

$$q_{ve}^2 = 4\pi^2 \varepsilon_v \Omega^2 D^4 = (2\pi \rho_0 D^2)^2, \quad q_{ve} = 2\pi \rho_0 D^2 \cong m/l. \tag{39}$$

As can be seen, the acoustic charge in the MKSA system is proportional to the mass of the vortex core which is the source of the field created by the vortex. Again the numerical constant 2 appears and we have not been able to remove it. Between the two acoustic charges, written in different unit systems, there is the relationship

$$\frac{q_v^2}{4\pi \varepsilon_v} \cong e_{vetN}^2, \tag{40}$$

similar to that for the electrostatic charge [40].

### 3.2. The angular momentum of the vortex

The specific quantity of the vortex is the angular momentum:

$$\vec{S}_v = \hat{z}\left[\int_0^D (\rho_0 v_{vi} r) 2\pi l r dr + \int_D^R (\rho_0 v_{ve} r) 2\pi l r dr\right] = \hat{z} 2\pi l \rho_0 \Omega \left[\int_0^D r^3 dr + D^2 \int_D^R r dr\right] = \pi D^2 l \rho_0 \left(R^2 - \frac{D^2}{2}\right) \Omega \hat{z} = m\left(R^2 - \frac{D^2}{2}\right) \Omega \hat{z}, \tag{41}$$

with $R$ radius of the enclosure.

For a system of two interacting vortices, to this intrinsic/internal angular momentum is added the orbital angular momentum $\vec{L}_{sv}$. If the vortices are identical, it is:

$$\vec{L}_{sv} = 2\hat{z}(\pi D^2 l \rho_0)\left(\frac{r}{2}\right) v_{ve} = (\pi D^4 l \rho_0) \Omega \hat{z} \tag{42}$$

and the total angular momentum of the two-vortex system is

$$\vec{L}_{svt} = 2\vec{S}_v + \vec{L}_{sv} = 2m\left(R^2 - \frac{D^2}{2}\right)\Omega \hat{z} + mD^2 \Omega \hat{z} = 2mR^2 \Omega \hat{z}, \tag{43}$$

if $r \gg 2D$.

### 3.3. Energy densities of the vortex field

To obtain the energy density expression of the vortex field, we substitute Eqs. (24) in Eq. (21)

$$w_i = \frac{\vec{E}_{vi}^2}{8\pi} + \frac{\vec{H}_{vi}^2}{8\pi} = \frac{\Omega^4 r^2}{8\pi} + \frac{\Omega^2}{8\pi}. \tag{44}$$

To express them in the international system (SI), $\langle w_{iSI} \rangle = Jm^{-3} = kg m^{-1} s^{-2}$, we use the relations:

$$w_{iSI} = \frac{(4\pi \varepsilon_v) \vec{E}_{vi}^2}{8\pi} + \left(\frac{4\pi}{\mu_v}\right)\frac{\vec{H}_{vi}^2}{8\pi} = \frac{\varepsilon_v \vec{E}_{vi}^2}{2} + \frac{\vec{H}_{vi}^2}{\mu_v 2} = \frac{\rho_0 (\Omega r)^2}{2} + \frac{\rho_0 u_0^2}{2}, \tag{45}$$

that is, we obtain the sum of the kinetic energy density and the relativistic-acoustic energy density of the fluid ($\rho_0 u_0^2 \simeq \zeta p_0$) inside the vortex core.

For the outside of the vortex, using Eqs. (25), it follows:

$$w_e = \frac{\vec{E}_{ve}^2}{8\pi} = \frac{\Omega^4 D^8}{8\pi r^6}, \ w_{eSI} = \frac{(4\pi\varepsilon_v)\vec{E}_{ve}^2}{8\pi} = \frac{\rho_0 \Omega^2 D^8}{2r^6} = \frac{\rho_0 \Omega^2 D^4}{2r^2}\frac{D^4}{r^4}, \qquad (46)$$

which is lower than the acoustic kinetic energy density outside the vortex core $w_{vka} = (\rho_0/2)(\Omega D^2/r)^2$.

Maxwell's hydrodynamic equations are most similar to the equations of the gravito-electromagnetic field [41-44]. For the weak gravitational field, the gravitational charge density is the mass density $\rho_g = \rho = dm/dV$, and the Maxwell equation for the gravitational field strength is $\nabla \cdot \vec{E}_g = 4\pi G \rho = \rho/\varepsilon_g$, $\varepsilon_g = 1/(4\pi G)$. It is interesting that this gravitational permittivity, which is a function of the gravitational constant $G$ can be put in the form analogous to that for the hydrodynamic/acoustic permittivity. For a universe model of finite radius $R_u$ and density $\rho_u$, oscillating with frequency $H_u = 1/T_u = c/R_u$ and/or rotating with the same oscillation period, according to relativistic universe models [45 - Ch. 16, Eq. (16.3.8)], between these quantities there is the relationship $\rho_u \cong H_u^2/G$ and so $\varepsilon_g \cong 1/G = \rho_u/H_u^2$.

In the same relativistic universe, at the submicroscopic scale, i.e. at the Planck scale [46], the Planck density $\rho_P \cong m_P/l_P^3$, $m_P = \sqrt{\hbar c/G}$ and the Planck pulsation $\omega_P = m_P c^2/\hbar$ $G \cong H_u^2/\rho_u$ are defined, so that $G \cong \omega_P^2/\rho_P$ and therefore $\varepsilon_g \cong 1/G = \rho_P/\omega_P^2$. These similarities highlight the role of the background field (the fundamental fluid with periodic movements – oscillating, in the case of pulsating bubbles [15], and rotating for vortices) in the generation of interactions between systems specific to the acoustic world and the possibility of deriving hydrodynamic Maxwell equations.

## 4. The vortex as an acoustic lens

### 4.1. Acoustic index of refraction around a vortex

The velocity field, specific to the vortex, $\vec{v}_v$, generates in a compressible fluid density variations, $\rho_v = \rho_0 v_v/u_0$, and pressure variations, $p_v = \rho_0 u_0 v_v$, [39- Ch. 8] which locally modify the propagation speed of the acoustic waves and therefore a refractive index of the fluid, $n_v$, in the presence of a vortex, According to the works [5, 6, 11], the refractive index for the vortex has the expressions:

$$n_v(\rho) = \left(1 + \frac{\rho_v}{\rho_0}\right)^{-\zeta} = \left(1 + \frac{v_v}{u_0}\right)^{-\zeta} \cong 1 - \frac{\zeta v_v}{u_0} + \frac{\zeta(\zeta+1)}{2}\left(\frac{v_v}{u_0}\right)^2, \qquad (47)$$

$$n_v(p) = \left(1 + \frac{p_v}{p_0}\right)^{\frac{-\zeta}{2\zeta+1}} = \left(1 + \frac{\rho_0 u_0 v_v}{p_0}\right)^{\frac{-\zeta}{2\zeta+1}} \cong 1 - \frac{\zeta}{2\zeta+1}\frac{\rho_0 u_0 v_v}{p_0} + \frac{\zeta(3\zeta+1)}{2(2\zeta+1)^2}\left(\frac{\rho_0 u_0 v_v}{p_0}\right)^2. \qquad (48)$$

Since $p_0 = \rho_0 u_0^2/(2\zeta+1)$, substituting in Eq. (45), it follows

$$n_v(p) \cong 1 - \frac{\zeta v_v}{u_0} + \frac{\zeta(3\zeta+1)}{2}\left(\frac{v_v}{u_0}\right)^2, \qquad (49)$$

which differs from the index expression (47) by the second power of the ratio $v_v/u_0$. The two expressions of the index of refraction are different, even if we average over time (as in the case of the field of an acoustic wave [5, 6]), for a velocity field in the vortex that does not depend on time, because $\langle v_v \rangle \neq 0$. For a better approximation of the refractive index expression, in the following we will use the expression of the average of the two Eqs. (47, 48)

$$n_\upsilon(\upsilon_\upsilon) = \frac{n_\upsilon(\rho) + n_\upsilon(p)}{2} \cong 1 - \frac{\zeta \upsilon_\upsilon}{u_0} + \frac{\zeta(2\zeta+1)}{2}\left(\frac{\upsilon_\upsilon}{u_0}\right)^2. \tag{50}$$

For $r < D$ and $\upsilon_{\upsilon i} = \Omega r$, the index of refraction inside the nucleus/ core is:

$$n_{\upsilon i} = 1 - \frac{\zeta \Omega r}{u_0} + \frac{\zeta(2\zeta+1)}{2}\left(\frac{\Omega r}{u_0}\right)^2 \tag{51}$$

and for $r > D$ and $\upsilon_{\upsilon e} = (\Omega D^2)/r$, the refractive index outside the core is

$$n_{\upsilon e} = 1 - \frac{\zeta \Omega D^2}{u_0 r} + \frac{\zeta(2\zeta+1)}{2}\left(\frac{\Omega D^2}{u_0 r}\right)^2. \tag{52}$$

For velocities in the vortex much lower than the speed of sound in the respective fluid, $\Omega D \ll u_0$, the refractive indices can be approximated by the relations: $n_{\upsilon i} = 1 - \zeta \Omega r/u_0 < 1$ and $n_{\upsilon e} = 1 - \zeta \Omega D^2/(u_0 r) < 1$.

### 4.2. Acoustic lens generated by a vortex

The existence of an acoustic refractive index around a vortex causes the vortex to behave like a cylindrical acoustic lens with focal length [5, 6]:

$$\frac{1}{f_\upsilon} = \frac{2(n_\upsilon - 1)}{r} = \frac{2\zeta \upsilon_\upsilon}{u_0 r}\left[\frac{(2\zeta+1)\upsilon_\upsilon}{2u_0} - 1\right] \tag{53}$$

or

$$f_\upsilon = \frac{u_0 r}{2\zeta \upsilon_\upsilon\left[\frac{(2\zeta+1)\upsilon_\upsilon}{2u_0} - 1\right]}. \tag{54}$$

The lens corresponding to the vortex is converging, $f_\upsilon > 0$, if $\upsilon_\upsilon/u_0 > 2/(2\zeta+1)$ and is divergent if $\upsilon_\upsilon/u_0 < 2/(2\zeta+1)$. As en example, for water [5, 6] $\zeta = 3$ and $\upsilon_\upsilon/u_0 > 2/7$, that is, the lens is convergent at speeds close to the speed of sound in water. At low velocities in the vortex, the lens is divergent, that is, it scatters the acoustic waves. If $\upsilon_\upsilon/u_0 = 2/(2\zeta+1)$, (and for points on the vortex axis, $\upsilon_\upsilon = 0$ and $f_\upsilon \to \infty$) the lens has infinite focal length and the waves pass undeviated.

### 4.3. Dumb hole corresponding to the vortex

The acoustic lens becomes a dumb hole (an acoustic black hole) if the acoustic wave (acoustic ray) is captured [5, 6] i.e. it is deflected on a circular or spiral path intersecting the axis of the vortex. In this situation the focal length is equal to or less than the impact distance, $f_\upsilon \le r = r_{\upsilon DH}$. With this condition, the equation that determines the radius of the dumb hole is obtained, $r_{\upsilon DH}$,

$$2\zeta \upsilon_\upsilon(r_{\upsilon DH})\left[\frac{(2\zeta+1)\upsilon_\upsilon(r_{\upsilon DH})}{2u_0} - 1\right] = 1. \tag{55}$$

With this equation, for the exterior of the core vortex, $r > D$, $\upsilon_{\upsilon e} = (\Omega D^2)/r$, we obtain the equation:

$$\frac{2\zeta \Omega D^2}{u_0 r_{\upsilon DHe}}\left[\frac{(2\zeta+1)\Omega D^2}{2u_0 r_{\upsilon DHe}} - 1\right] = 1, \; r_{\upsilon DHe}^2 + \frac{2\zeta \Omega D^2}{u_0} r_{\upsilon DHe} - \frac{\zeta(2\zeta+1)\Omega^2 D^4}{u_0^2} = 0. \tag{56}$$

This equation has as a real solution the expression:

$$r_{vDHe} = \frac{\Omega D^2 \left(1 + \sqrt{1 + (2\zeta + 1)/\zeta}\right)}{u_0} \tag{57}$$

and this radius is smaller than the core radius, $r_{vDHe} < D$, if $\Omega D \left(1 + \sqrt{1 + (2\zeta + 1)/\zeta}\right) < u_0$.

For the vortex core, $r < D$, $v_{vi} = \Omega r$, the equation that determines the radius of the dumb hole is

$$\frac{\zeta(2\zeta + 1)\Omega^2 r_{vDHi}^2}{u_0^2} - \frac{2\zeta \Omega r_{vDHi}}{u_0} - 1 = 0. \tag{58}$$

This equation has as a real solution the expression:

$$r_{vDHi} = \frac{u_0 \left(1 + \sqrt{1 + (2\zeta + 1)/\zeta}\right)}{(2\zeta + 1)\Omega} \tag{59}$$

and depends on the constant $\zeta$ that characterizes the phenomenon of fluid compression, the speed of sound in the undisturbed fluid $u_0$ and the angular velocity $\Omega$ of the fluid in the core of the vortex.

## 5. Conclusions

In this paper we have demonstrated, using Maxwell's hydrodynamic equations, that the vortex, as a system having internal angular momentum, $\vec{S}_v \neq 0$, from the point of view of interaction with other vortices, behaves like an acoustic dipole, if the distance between them is $r > D$.

As for the pulsating bubble (acoustic monopole), the hydrodynamic equations for the vortex show an acoustic permittivity, $\varepsilon_v = \rho_0/\Omega^2$, which depends on the undisturbed fluid density and angular velocity. In this way, the role of the fluid that has periodic properties in the generation of interactions specific to the AW is highlighted. These periodic properties of the acoustic background explain the boundary conditions imposed by the quantification of the characteristic properties of the systems specific to the AW and, by analogy, to the Electromagnetic World [47, 48]. It remains to be studied, in a future work, the oscillations of the vortex induced by an inducing acoustic wave and the particular state when the pulsation of the acoustic wave is equal to the rotation speed of the vortex, $\omega = \Omega$.

We have also demonstrated that the energy density of the electric acoustic field of the vortex core is the rotational kinetic energy density, $w_{iESI} = \rho_0 (\Omega r)^2 / 2$ and the energy density of the magnetic component is proportional to the fluid pressure, $w_{iHSI} = \rho_0 u_0^2 / 2$. Also, since the vortex induces changes in the fluid density, $\rho_v = \rho_0 v_v / u_0$, and pressures, $p_v = \rho_0 u_0 v_v$, and therefore in the propagation speed of the acoustic waves, the corresponding acoustic refractive index causes the vortex to behave like an acoustic lens. We evaluated this acoustic refractive index and the possibility of the vortex becoming a dumb hole. These results allow solving the problems: pulsating vortex under the action of an acoustic wave, the multi-vortex system, and the pulsating bubble system captured by a vortex (i.e., the model of the monopolar acoustic particle with spin).